\documentclass[twocolumn]{aastex62}

\usepackage{comment}
\usepackage{graphicx}  
\usepackage{multirow}

\graphicspath{{./}{figures/}}

\shortauthors{Smith et al.}

\emergencystretch=\maxdimen
\begin{document}


\title{Probing Binary Black Hole Formation Channels through Cosmic Large-Scale Structure}

\correspondingauthor{William J. Smith}
\email{william.j.smith@vanderbilt.edu}

\author[0009-0003-7949-4911]{William J. Smith}
\affiliation{Department of Physics and Astronomy, Vanderbilt University, 2301 Stevenson Center, Nashville, TN 37235}

\author{Krystal Ruiz-Rocha}
\affiliation{Department of Physics and Astronomy, Vanderbilt University, 2301 Stevenson Center, Nashville, TN 37235}

\author{Kelly Holley-Bockelmann}
\affiliation{Department of Physics and Astronomy, Vanderbilt University, 2301 Stevenson Center, Nashville, TN 37235}
\affiliation
{Department of Physics, Fisk University, Nashville, TN 37208}

\author[0000-0001-8799-2548]{Michela Mapelli}
\affiliation{Universit\"at Heidelberg, Zentrum f\"ur Astronomie (ZAH), Institut f\"ur Theoretische Astrophysik, Albert-Ueberle-Str. 2, 69120, Heidelberg, Germany}

\affiliation{Universit\"at Heidelberg, Interdisziplin\"ares Zentrum f\"ur Wissenschaftliches Rechnen, Heidelberg, Germany}

\author[0000-0003-1007-8912]{Karan Jani}
\affiliation{Department of Physics and Astronomy, Vanderbilt University, 2301 Stevenson Center, Nashville, TN 37235}




\begin{abstract}
The growing number of binary black hole mergers detected through gravitational waves offers unprecedented insight into their underlying population, yet their astrophysical formation channels remain unresolved. We present a new method to distinguish binary black hole formation channels using their spatial clustering at cosmological scales. Employing the cosmological hydrodynamic simulation \textit{Illustris}, we trace the distribution of mergers across cosmic time and compare them with the underlying matter distribution associated with three candidate origins: isolated binary stellar evolution, binaries embedded in AGN disks, and primordial black holes within dark matter halos. For mergers at redshift $z \lesssim 0.5$, these channels show distinct clustering signatures that could be accessible with proposed upgrades to current ground-based gravitational-wave detectors. Using mock catalogs for next-generation facilities such as Cosmic Explorer, we find that their sensitivities would enable differentiation of these formation pathways out to redshift $z \sim 5$ within the first decade of observations. This approach provides a new framework to link gravitational-wave populations with the large-scale structure of the Universe. By treating black hole mergers as cosmological tracers, our results demonstrate how cross-correlations between gravitational-wave catalogs and the cosmic matter field can constrain the relative contribution of stellar, AGN, and primordial channels, offering a complementary probe to population-inference studies. These findings underscore the emerging potential of gravitational-wave cosmology to reveal where and how black holes form and merge across cosmic history.

\end{abstract}




\section{Introduction} \label{sec:intro}

Research into gravitational wave (GW) stellar mass binary black hole mergers (sBBHs) has expanded dramatically since the detection of GW150914 nearly a decade ago ~\citep{Abbott_2016}. Today's LIGO-Virgo-KAGRA (LVK) collaboration can ideally detect events nearly daily through the Advanced LIGO ~\citep{Advanced_ligo_2015}, Advanced Virgo ~\citep{Acernese_2014}, and KAGRA ~\citep{akutsu2020_KAGRA} detectors, resulting in nearly one hundred detections in the first through fourth observing runs ~\citep{GWTC_1, GWTC_2, GWTC_2p1, GWTC_3} and many more with the newly released GWTC-4.0 catalog ~\citep{GWTC_4}. These detections have resulted in the first generation of observational evidence about the population of sBBHs and the underlying astrophysical phenomena which give rise to this population ~\citep{compact_binaries_pop_o3}. Future ground-based detector networks, including advancements in the LVK network at A+ sensitivity ~\citep{LVK_obs_plan}, and next-generation facilities like Cosmic Explorer (CE) ~\citep{reitze2019_CE} and Einstein Telescope (ET) ~\citep{Punturo_2010_ET, ET}, will be able to detect multitudes of sources ~\citep{Ng_2021, maggiore2024, Future_detector_localization}. Additionally, the Laser Intererometer Space Antenna (LISA) is expected to detect these sources at lower frequencies for multiband observations ~\citep{Jani_2019, LISA_redpaper, ruizrocha2024, Ranjan_multiband}.

One of the most prominent open questions for understanding the sBBH population is that of their formation channels ~\citep{Gerosa_2021, Rates_of_CBCs, Li_2025, banagiri_2025a}, the most commonly considered of which are the stellar binary channel ~\citep{Bethe_1998, Belczynski_2002, Mennekens_2014, Belczynski_2016, de_Mink_2016, Marchant_2016, Giacobbo_2017, Kruckow_2018}, the dynamical channel ~\citep{Banerjee_2010, Ziosi_2014, Rodriguez_2015, Askar_2016, Rodriguez_2016b, Banerjee_2017, Rodriguez_2021, Fragione_2023, banagiri_2025b}, and the active galactic nucleus (AGN) disk channel ~\citep{AGN_formation, Gayathri_2023, Vaccaro_2024,  zhu_2025}. Another proposed formation channel is that some sBBH mergers are primordial in origin ~\citep{Clesse_2020}.  Although much research has been done on the stellar binary evolution and dynamic channel, the AGN channel is exciting because of its potential to explain sBBH observations with masses in the pair instability supernova (PISN) mass gap ~\citep{McKernan_2012_AGN_IMBH}. Much research has gone into understanding the complex physical processes that result in sBBH mergers in AGN disks ~\citep{Stone_2016, Graham_2020, Secunda_2019, Li_2021, Ford_2022, Li_2022, Rowan_2023, Ishibashi_2024, Rom_2024}. 

In addition to the proposals of formation channels that could result in sBBH mergers themselves, much work has gone into developing methods that can differentiate them based on observational evidence and answer the question of which of the proposed formation channels gives rise to which subsets of the observed sBBH population ~\citep{Vitale_2017, Zevin_2021, Abbott_2023}. Some tests of these formation channels include analyzing the merger rate to redshift relationship ~\citep{Fishbach_2018, Fragione_2018,  Yang_2020, Santoliquido_2020, Mapelli_2022, van_Son_2022}, the sBBH binary separation ~\citep{Jani_2020_Budget}, the sBBH mass ratio distribution ~\citep{Fishbach_2020}, the sBBH spin distribution ~\citep{Rodriguez_2016, Farr_2017,Stevenson_2017, Sedda_2020_spin, Biscoveanu_2022}, the sBBH orbit eccentricities ~\citep{Samsing_2017, ligo_eccentricity2019, Romero_Shaw_2020, Romero_Shaw_2022}, and, in the case of AGN, direct multimessenger observation ~\citep{Graham_2023, Tagawa_2023}. Additionally, some works have explored using underlying astrophysics as a means to probe sBBH rates resulting from the AGN formation channel ~\citep{Bartos_2017, McKernan_2022, Veronesi_2022, vaccaro_2023, mckernan_2023} or from primordial black holes ~\citep{Ali_Ha_moud_2017, Scelfo_2018, Chen_2020, Hall_2020, De_Luca_2021, Franciolini_2022, raidal_2024, bouhaddouti_2025}.

Another approach to studying the origins of sBBH is quantifying the bias parameter between their mergers with host galaxies ~\citep{Banagiri_2020, Vijaykumar_2024, Dehghani_2025} and 
underlying matter distributions ~\citep{Libanore_2021}. These bias parameters describe the relationships between two spatial clustering distributions. The bias between sBBHs and their galaxy distribution can help inform our knowledge of how and where sBBHs are likely to arise ~\citep{Mapelli_2018_host_galaxy, Scelfo_2020}, and the bias between sBBHs and the dark matter distribution will provide a new tool to map the dark matter distribution in the universe given the non-direct observability of dark matter ~\citep{Mukherjee_biases}. Understanding and constraining the sBBH to galaxy bias parameter is also an important component of cosmological techniques like the cross-correlation technique for constraining the Hubble constant ~\citep{Mukherjee_H0, afroz_2024}. A few works have explored correlation techniques for both the AGN channel ~\citep{Bartos_2017b, Veronesi_2022, Veronesi_2024} and PBH channel ~\citep{Atal_2020, Libonare_2023}, as well as galaxy-SMBH cross-correlation ~\citep{Sah_2024}. The main benefit of correlation techniques is that they only rely on spatial clustering, not intrinsic properties of sBBH events, which makes available new techniques to test them that are not feasible with other methods of testing formation channels. 

In this paper, we propose using the clustering bias of sBBHs in a new way - as a novel method to distinguish sBBH formation channels. Additionally, we use data from the hydrodynamic simulation Illustris-1 (hereafter referred to as `Illustris') in an innovative way to provide a first test of this method. We carry out the first investigation of correlation techniques for the stellar binary channel, the AGN channel, and the PBH channel using a hydrodynamic simulation. This allows us to not only analyze the clustering signatures, but to do so in a single testbed environment.

We employ the cosmological hydrodynamic simulation \textit{Illustris} as our testbed to evaluate our method, generating synthetic observations of stellar mass mergers assuming different origin channels to investigate the clustering of simulated sBBHs (e.g., ~\citealt{peron2023clusteringbinaryblackhole}), the goal of this paper is to explore how well clustering statistics can be used in gravitational wave catalogs (observations) to probe astrophysical formation channels. 

Our application of this method leverages the fact that by using a hydrodynamic simulation, we know a priori the formation channel of our simulated data set. This offers two advantages. First, it allows us to calculate a clustering bias between the sBBHs and the stars that form them, assuming the sBBHs are formed solely through the stellar binary channel, and second, it allows us to calculate the clustering bias between sBBHs and AGN, and the sBBHs and dark matter particles, assuming the sBBHs are formed through the stellar channel. 

In Section \ref{sec:data}, we introduce the Illustris dataset and the stellar binary population synthesis code to seed sBBHs through the cosmological volume based on age and metallicity of the stellar particles ~\citep{Mapelli_2017}. In Section \ref{sec:GW_detection} we explain our procedure for creating a synthetic mock-observed catalog in a future CE-like detector using the method of ~\citet{Chen_2022} and applied by ~\citet{ruizrocha2024}, and how we use this to calculate clustering biases for the synthetic mock-observed catalog. In Section \ref{sec:2pcf}, we explain our methods for calculating the two-point correlation functions, relative clustering biases, and the resampling technique we use to calculate confidence intervals for each of these biases to see if the biases are distinguishable from each other when the true source of sBBH mergers is assumed in the mock-synthetic catalog. In Section \ref{sec:results} we report a snapshot `census' of our data, including the simulated detection rate at multiple redshifts, the two-point correlation functions and relative clustering biases at those redshifts, and our analysis of the evolution of the relative clustering biases over a ten year observing period for CE and A+ detector sensitivities using the mock-observed catalog. Finally, in Section \ref{sec:Discussion}, we discuss the assumptions and limitations of these results, as well as implications for future work determining formation channels with this method.

\textcolor{red}{ }


\section{Methods} \label{sec:methods}

\subsection{The Illustris Simulation and the sBBH Dataset} \label{sec:data}

Illustris is a large-scale hydrodynamic simulation with a comoving box size of $(106.5 \ Mpc)^3$ \citep{Vogelsberger_2014,Nelson_2015}. It comprises 135 snapshots from $z = 127$ to $z = 0$ and is evolved using the \textsc{Arepo} moving mesh code with cosmological parameters from WMAP-9 ($h = 0.704$, $\Omega_{m} = 0.2726$, $\Omega_{\Lambda} = 0.7274$, $\Omega_{b} = 0.456$, and $\sigma_{8} = 0.809$). Illustris has a baryonic particle resolution of $1.26 \times 10^{6}$ M$_\odot$ and a dark matter particle resolution of $6.26 \times 10^{6}$ M$_\odot$, along with a multitude of sub-grid physics prescriptions (i.e., cooling, star formation, AGN feedback, supermassive black hole formation and accretion, among others). Using the Illustris public release API (\url{https://www.illustris-project.org}), we download the stellar particles and associated information in each snapshot, including their unique particle identification numbers and coordinates in the snapshot. We additionally download dark matter, gas, and black hole particle data, including particle ID and spatial coordinates.

In addition to Illustris data, we also use data from ~\citet{Mapelli_2017}, who introduced a method of seeding Illustris with  sBBHs (hereafter referred to as the Illustris-sBBH dataset). They utilized a population synthesis code to evolve stellar binaries, followed by a Monte Carlo approach to assign sBBHs to stellar particles in Illustris. This assignment was based primarily on stellar particle metallicities in Illustris and the binary population synthesis models. Consequently, each stellar particle in Illustris was associated with a set of BBH mergers, each with a specific look-back time. They evaluated six different population synthesis models, varying prescriptions for the common envelope phase, SN properties, and natal kick treatment, among others. For this dataset, the upper mass limit of the primary black hole is $\lesssim 40 \ M_{\odot}$. In this preliminary study, we ignore the emerging population of black holes in the PISN mass-gap ~\citep{LIGO_gwtc4, GWTC4_pop} and the detections of `lite'-intermediate-mass black hole range ~\citep{Udall_2020, Ruiz-Rocha_2025, gw_231123}. 

Based on our mass criteria, we adopt the ``DK'' model from Illustris-sBBH ~\citep{Mapelli_2017} because it has the closest consistency to the sBBH cosmic merger rate reported in the Gravitational-wave Transient Catalog (GWTC-3) results from LVK ~\citep{compact_binaries_pop_o3}, and further consistent with the recent GWTC-4 release ~\citep{GWTC4_pop}. This model assumes natal kicks for BBHs that are as large as the natal kicks of young pulsars observed in the Milky Way (~\citealt{Hobbs2005}; see also ~\citealt{Sgalletta2025} and ~\citealt{Boesky2024} for a discussion of this assumption).

We combine the Illustris and Illustris-BBH data using the stellar particle ID to join the location data of each stellar particle in Illustris associated with a merger in the Illustris-BBH dataset. To do this in practice, we first choose the redshift of interest. Because the Illustris-BBH data is continuous and the Illustris data is contained in snapshots, we adopt a $\delta_{t}$ for the look-back time in the Illustris-BBH data to constrain the merger data to the data comparable with the Illustris snapshot. For example, for $z = 1$, we first find the Illustris snapshot associated with $z = 1$. Then, to merge the particle location data in the snapshot to the continuous Illustris-BBH data, we choose a $\delta_{t}$. We use $\delta_{t} = 0.0001 \ Gyr$ for this work. We then identify the subset of mergers in the Illustris-BBH data corresponding to the redshift of the Illustris snapshot closest to $z = 1$ and $z = 1 + 0.0001 \ Gyr$, and consider those the mergers taking place in the snapshot. This allows us to identify the coordinates of each merger that can reasonably be associated with a particular snapshot.

\subsection{Detactability with ground-based GW detectors} \label{sec:GW_detection}

Next, we turn to what future detectors like CE and A+ will observe at a redshift of interest by taking the total number of sBBH mergers in the Illustris-BBH dataset from the redshift of interest $\pm 0.1$ (for example, at $z=1$, we use all sBBH mergers between $z=0.9$ and $z=1.1$). We calculate an estimated signal-to-noise ratio (SNR) using the look-back time and masses of mergers matched to the Illustris snapshot. Given the total number of sBBH mergers in this restricted range is still on order millions, calculating an SNR for so many sources is computationally unfeasible, we take a random subset of 1000 sBBH points in the redshift range and calculate the SNRs of each merger in that subset. From this SNR distribution of the subset, we can infer the total number of sources in the entire set above a specific SNR, in this work, we consider observations SNR$>8$. For each merger in the subset, we use the Python package \textsc{pycbc} ~\citep{Py_cbc} to generate a waveform for a source with given masses and distance (assuming Illustris cosmology). We use the `IMRPhenomXP' approximant and calculate two SNRs for each event, one using the LIGO Advanced Design T1800042 (A+) and the other using CE Design Sensitivity P160014 ~\citep{Srivastava_2022_CE_sensitivity}. For the SNR calculation, we additionally use a random uniform distribution from $-1$ to $1$ to assign an inclination angle through $|cos (\iota)|$ where $\iota$ is the inclination of the BBH system relative to the detector when calculating the SNR. The inferred number of BBH mergers with SNR$>8$ in each snapshot is reported in the `sBBH mergers' column of Table \ref{tab:particle_numbers}. We use this procedure to calculate the total number of detectable sources in the Illustris-sBBH data set within the redshift range. From this calculation, we can estimate the observation rate for a given detector characterization as follows:

\begin{equation}\label{observed_BBHs}
\frac{dN}{dz~dt} = \frac{d^{2}n(z)}{dzdV_{c}}\,{}\frac{dz}{dt}\,{}\frac{dV_{c}}{dz}\,{}\frac{1}{1+z},
\end{equation}

where $V_{c}$ is the co-moving volume, $n(z)$ is the number of BBHs at redshift $z$, $\frac{1}{1 + z}$ converts $\frac{dz}{dt}$ to the observer frame, and $\frac{d^{2}n(z)}{dzdV_{c}}$ is the rate found in the simulation volume above a certain SNR threshold, which can be expressed by

\begin{equation}\label{sim_conversion}
\frac{d^{2}n(z)}{dz~dV_{c}} = \frac{N(z)}{\Delta z ~ dV_{c}},
\end{equation}

where $N(Z)$ is the number of BBHs merging within a redshift bin, $\Delta z$ is the width of the redshift bin, and $V_{sim}$ is the simulation volume~\citep[see also][]{ruizrocha2024,Chen_2022}. For this work, we calculate the observed rate using 100 bins. The calculated rates for $z=0.5$, $z = 1$, $z= 3$ and $z = 5$ for both CE and A+ appear in the rightmost columns of Table \ref{tab:particle_numbers}. 

Next, we use the predicted 5-year merger rate for CE to select a subset of sBBH mergers matched to the Illustris snapshot at that redshift. To visualize the observed mergers, we display a slice of the simulation at that redshift (see the top row of Figure ~\ref{fig:5yr_big_multiplot}. Each slice is $1.2$ Mpc thick and contains a downsampled set of dark matter particles (green), stellar particles (orange), SMBHs (black), sBBH mergers (blue), and a subset representing the predicted number of observable mergers from CE over a five year observing run (red).

\subsection{Two point correlation and clustering bias} \label{sec:2pcf}

Next, we turn to measuring the clustering of each given particle type by calculating the two-point correlation function, $\xi(r)$, which is the excess probability of finding a pair of halos separated by a physical distance $r$, over what would be expected for a random distribution. To calculate $\xi(r)$, we use the equation

\begin{equation}\label{2pcf}
\xi(r) = \frac{DD(r)}{RR(r)} - 1,
\end{equation}

\noindent where $DD(r)$ is the number of halo pairs at separation distance $r$ in the data and $RR(r)$ is the number of pairs at separation distance $r$ in a random data set of equivalent volume. We use the \textsc{CorrFunc} package ~\citep{Corrfunc}, a suite of highly-optimized and robustly-tested codes for calculating clustering statistics.

\begin{table*}[ht!]
\centering
\begin{tabular}{cccccc}
\hline
\hline
\textbf{Redshift ($z$)} & \textbf{Stellar Particles} & \textbf{SMBHs} & \textbf{sBBH Mergers} & \textbf{CE: 1yr (5yr)} & \textbf{A+: 1yr (5yr)} \\
\hline
0.5 & $5.38 \times 10^{9}$ & $3.23 \times 10^{4}$ & $3.30 \times 10^{4}$ & 196 (980) & 115 (575) \\
1.0 & $5.48 \times 10^{9}$ & $3.07 \times 10^{4}$ & $6.10 \times 10^{4}$ & 579 (2895) & 85$^{*}$ (425) \\
3.0 & $5.77 \times 10^{9}$ & $1.74 \times 10^{4}$ & $1.29 \times 10^{5}$ & 813 (4065) & 0$^{*}$ (0) \\
5.0 & $5.98 \times 10^{9}$ & $4.51 \times 10^{3}$ & $7.24 \times 10^{5}$ & 235 (1175) & 0$^{*}$ (0) \\
\hline
\end{tabular}
\vspace{6pt}
\caption{\textbf{Illustris Snapshot Census by Redshift.} 
The table lists the number of stellar particles, supermassive black holes (SMBHs), and stellar-origin binary black hole (sBBH) mergers for selected redshift snapshots of the Illustris simulation. The number of detected events for CE and A+ network are for 1-year (5-year) observing periods. The asterisk ($^{*}$) indicates that A+ observed rates were calculated for $z=1$, $z=3$, and $z=5$, but are not used in this work due to low statistics.}
\label{tab:particle_numbers}
\end{table*}

\begin{figure*}[p]
    \centering
    \includegraphics[width = 0.93\textwidth]{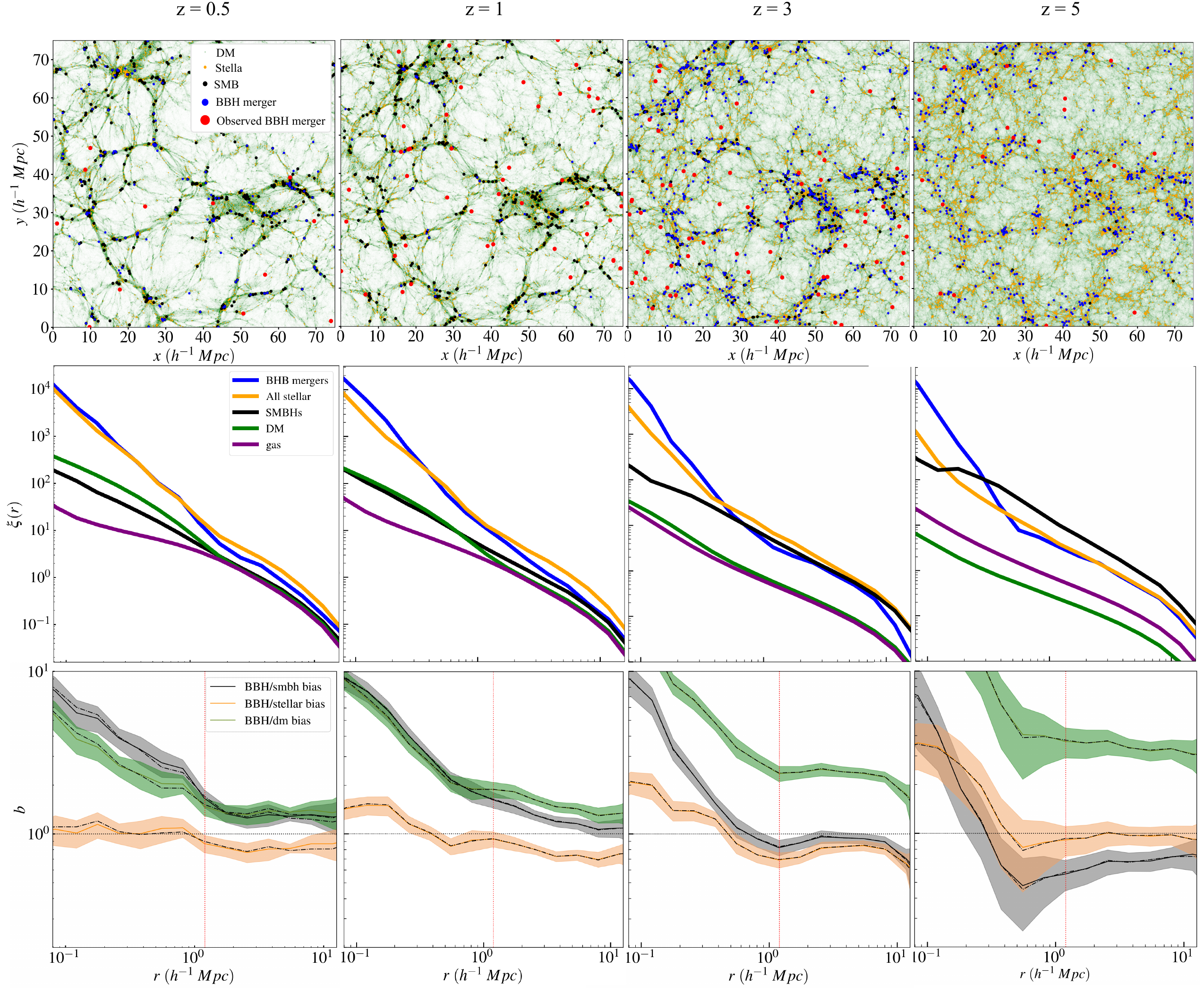}
    \caption{Visualization slices (top row), two-point correlation functions (middle row) and relative clustering biases (bottom row) for $z = 0.5$ (far left column), $z = 1$ (middle left column), $z = 3$ (middle right column), and $z = 5$ (far right column) data. Visualization slices include dark matter particles (green dots, representing a 10\% downsample of the total data), stellar particles (orange dots), SMBHs (black dots), all simulated sBBH mergers (blue dots), and sBBH mergers observed by CE in a five year observing run (red dots) for a slice thickness of $1.2$ Mpc. Two-point correlation functions include stellar particles (solid orange), SMBHs (solid black), dark matter particles (solid green), gas particles (solid purple), and all sBBH mergers in the Illustris-sBBH dataset matched to that redshift slice (solid blue). The stellar particles, dark matter particles, and gas particles are downsampled to a random sample of $10^{6}$ particles for computational feasibility. The relative clustering biases with 90\% confidence intervals include the mock-observed sBBH to SMBH bias (shaded purple), the mock-observed sBBH to stellar bias (shaded orange), and the mock-observed sBBH to dark matter bias (shaded gray). The dashed lines represent the true underlying bias of the simulated data, the dotted black horizontal line represents where no underlying bias is present, and the dotted red line represents the $1.2$ Mpc bin, used in creating Figure \ref{fig:bias_time}. The stellar clustering bias statistically significantly differs from the SMBH and DM bias at $z = 0.5$ and $z = 1$. The DM bias differs from SMBH and stellar at $z = 3$ and $z = 5$. Given CE sensitivity, the merger rate plays a larger part in constraining the relative clustering bias than redshift.}
    \label{fig:5yr_big_multiplot}
\end{figure*}

We calculate two-point correlation functions for stellar particles, dark matter particles, gas and SMBH particles in the snapshot. For computational expedience, we downsample the stellar, dark matter, and gas particles, choosing $10^{6}$ random particles of each type in the snapshot to allow enough particles for a robust statistic that is also computationally feasible. We also calculate the two-point correlation functions of all sBBH mergers associated with the snapshot. The two-point correlation functions for $z = 0.5$, $z = 1$, $z = 3$, and $z = 5$ are in the middle column of Figure~\ref{fig:5yr_big_multiplot}.

To determine the difference in clustering between the field binary stellar evolution channel, the AGN formation channel and primordial black holes, we use the coordinates of the sBBH, stellar particles, SMBH sources, and DM particles to calculate the relative clustering bias using the equation: 

\begin{equation}\label{bias}
b(r) = \sqrt{\frac{\xi(r)_{sBBH}}{\xi(r)_{particle}}},
\end{equation}

where $\xi(r)_{sBBH}$ is the two-point correlation function of a randomly-selected subset of sBBHs at the redshift above the SNR threshold, and $\xi(r)_{partcile}$ is the two-point correlation function of either stellar particles, SMBHs, or DM particles for the given redshift, with the assumption that we expect mergers within AGN disks and mergers of stellar binary origin to have the same SNRs at the same redshifts.

Errors for the observed relative clustering bias are calculated using resampling techniques on both $\xi(r)_{sBBH}$ and $\xi(r)_{particle}$. To calculate the error on $\xi(r)_{sBBH}$, one hundred distinct random subsets of sBBHs matching the CE predicted number are taken from all sBBH mergers at the redshift, and $\xi(r)_{sBBH}$ is calculated for each subset. From the samples, we can calculate the mean and standard deviation for each bin. We additionally inspect and verify the Gaussianity of the distribution of 100 sBBH draws on the separation distance we investigate. The errors for the clustering of SMBHs, stellar particles, and DM particles are calculated using bootstrapping with ten bootstrapped samples of 90\% of the particle type in the snapshot with replacement. From the error estimates for $\xi(r)_{sBBH}$ and $\xi(r)_{particle}$, we calculate the error on $b(r)$ using error propagation. We additionally note that this method results in significantly larger errors than the Poisson error for the same data.

\begin{figure*}[t!]
    \centering
    \includegraphics[width = 0.92\textwidth, height=0.95\textheight, keepaspectratio]{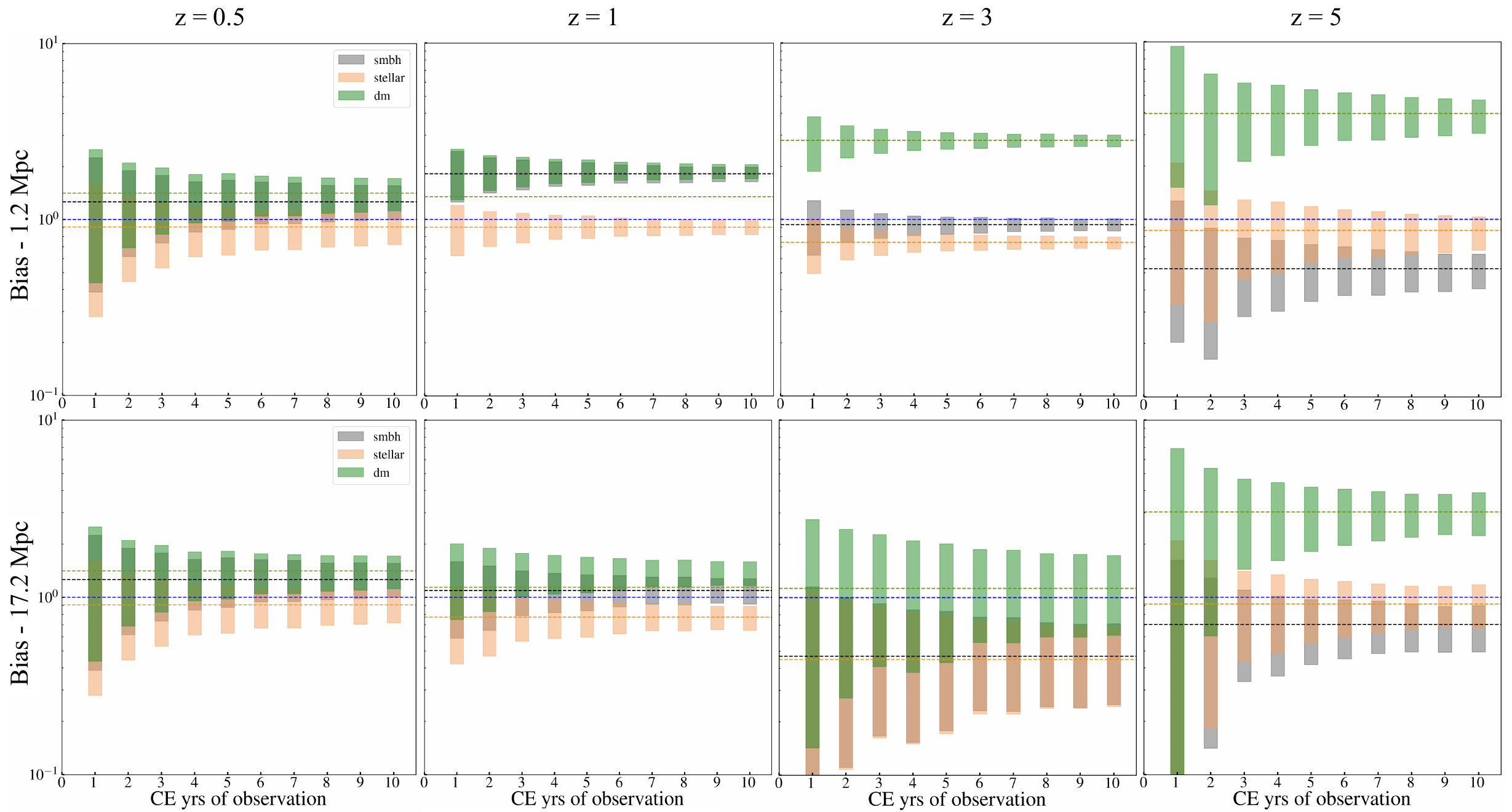}
    \caption{Evolution of 90\% confidence intervals of relative clustering biases of SMBH to observed sBBHs, stellar particles to observed BBHs, and dark matter particles to observed sBBHs, by years of observation on a CE-like detector at different redshifts ($z = 0.5$ far left column, $z = 1$ middle left column, $z = 3$ middle right column, and $z = 5$ far right column). The top row shows a separation distance of 1.2 Mpc (represented by the vertical dotted red lines on the bottom row of  \ref{fig:5yr_big_multiplot}, and the bottom row shows a separation distance of 17.2 Mpc. From the Figure, $z = 1$ at the closer separation distance shows by far the highest chance of non-overlap between the sBBH to stellar and sBBH to SMBH relative clustering biases, with marginal non-overlapping regions in $z = 3$ and $z = 5$ for 1.2 Mpc, and for $z =1$ at 17.2 Mpc.}
    \label{fig:bias_time}
\end{figure*}

Finally, we investigate the evolution of the bias error as a function of observing time at two distance scales, $1.2$ Mpc and $17.2$ Mpc, to determine whether additional observations over longer time spans will differentiate the clustering biases. We chose $1.2$ Mpc and $17.2$ Mpc because they are the midpoints of two of the logarithmic bins used in calculating the two point correlation function - $1.2$ Mpc is chosen because it is the the first bin larger than a Milky Way-like halo, and $17.2$ Mpc because it is the largest bin we used in calculating the two-point correlation function. Because these distance scales extend beyond the common radii for dark matter halos, a bias between sBBH merger clustering and stellar clustering would indicate that sBBH mergers happen preferentially in specific non-random subsets of galaxies, adding new insight to works that have explored the galactic hosts of sBBH mergers both from a simulation and observational perspective ~\citep{Toffano_2019_host_galaxies, Artale_2019, Aratale_2020, Adhikari_2020, Vijaykumar_2023}. Similarly, for AGN and PBH channels, these distances capture the effects of clustering on a scale slightly larger and much larger than most halos. We compute the bias calculation with observed sBBH rates in CE from one to ten years for $z = 0.5$, $z = 1$, $z = 3$, and $z = 5$ (Fig.~\ref{fig:bias_time}). The $1.2$ Mpc distance bin corresponds to the red dotted lines in the bias plots.

From this, we can show how the error shrinks with additional observations and constrain the minimum time necessary to differentiate an AGN disk formation channel from a stellar binary formation channel if the formation channel is entirely stellar in origin, based on the clustering of objects at this distance scale.

\section{Results} \label{sec:results}

\subsection{Simulation Particles}

Table ~\ref{tab:particle_numbers} shows a census of the particle type for each of the four redshifts analyzed. Though stellar particles and SMBHs  are taken from the simulation directly, the rest of the columns we calculate. As expected, the total number of sBBH mergers in each snapshot and the number of mergers that CE observes both peak at $z = 3$. We also note that, although $z = 5$ has more total mergers than $z = 1$, more are observed at $z = 1$ than $z = 5$. The first row of Figure ~\ref{fig:5yr_big_multiplot} shows the visualization slices of the sBBHs, SMBHs, and the synthetic mock observed catalog of sBBH mergers in the Illustris $z = 0.5$, $z = 1$, $z = 3$, and $z = 5$ snapshots. The evolution of structure through each redshift is apparent, especially with the dark matter and stellar particles, but also somewhat with the SMBHs and sBBH mergers, and not necessarily for the observed sBBHs.

\subsection{Two Point Correlation Functions}

The second row of Figure ~\ref{fig:5yr_big_multiplot} shows the two-point correlation functions of Illustris stellar, SMBH, gas, and dark matter particles at $z = 0.5$, $z = 1$, $z = 3$, and $z = 5$, along with the two-point correlation functions of all sBBH mergers matched to that Illustris snapshot. As expected, the clustering of the sBBH mergers tends to most closely match that of the stellar particles, with some minor but notable deviations. Of particular note is how the clustering of the sBBHs compared to the stellar particles shows different behaviors at smaller distances (below roughly 1 Mpc) and larger distances (above roughly 1 Mpc). At larger distances, the clustering of the sBBH mergers is slightly lower than the overall stellar correlation function, with the exception of $z = 5$, where they closely align. At smaller distances, the sBBH merger clustering is greater, an effect that is most significant at high redshift and decreases to lower redshifts. Additionally, the sBBH mergers show a noticeable change to a steeper slope at $z = 5$ and less pronounced at $z = 3$ below about 1 Mpc not seen in the other particle types.

When compared to the SMBH clustering, the sBBH and stellar clustering show a marked relative shift. At $z = 5$, the SMBH clustering is greater than the stellar particles and sBBHs for most distances. At $z = 3$, they are nearly identical at larger distances, until, at smaller distances, the sBBH and stellar clustering increase, and at $z = 1$ and $z = 0.5$, the stellar and sBBH clustering is consistently greater than the SMBH clustering across all distance scales. Next, we discuss how these changes in the two-point correlation functions manifest in the bias.

\begin{table*}[t!]
\centering
\begin{tabular}{cc|cccc}
\hline
\hline
\textbf{Separation Distance} & \textbf{Bias Comparison} & \textbf{$z = 0.5$} & \textbf{$z = 1$} & \textbf{$z = 3$} & \textbf{$z = 5$} \\
\hline
\multirow{2}{*}{1.2 Mpc}  & stellar/SMBH & 2 yrs & 1 yr & 6 yrs & 9 yrs \\
                          & stellar/DM   & 3 yrs & 1 yr & 1 yr & 3 yrs \\ \\
\multirow{2}{*}{17.2 Mpc} & stellar/SMBH & $>$10 yrs & 8 yrs & $>$10 yrs & $>$10 yrs \\
                          & stellar/DM   & $>$10 yrs & 4 yrs & $>$10 yrs & 3 yrs \\
\hline
\end{tabular}
\caption{\textbf{Years of CE Observation Until Differentiable Channels at Selected Clustering Separation Distances.}
Shown are the years of Cosmic Explorer (CE) observation required to statistically differentiate between stellar-origin and supermassive black hole (SMBH) or dark matter (DM) clustering channels, at representative redshifts. Longer timescales ($>$10 yrs) indicate degeneracy within the simulated signal population.}
\label{tab:yrs_table}
\end{table*}

\subsection{Relative Clustering Biases}

From the bias figures (bottom row of Figures ~\ref{fig:5yr_big_multiplot}), we first note that, as expected, the sBBH merger to stellar bias is closer to 1 (no bias) than any other particle type. However, even more noteworthy is that the bias is confidently less than one for greater distances with the exception of $z = 5$, and at smaller distances, the bias becomes quickly positive at higher redshifts but is least pronounced at lower redshifts. For $z = 0.5$ with a five-year observing rate, we see that the bias between stellar particles and sBBH mergers is nearly one (no bias) across all distance scales, but critically, at around the 1 Mpc distance scale and greater, the value falls just outside 90\% credible interval. Also, as expected, the sBBH merger to dark matter bias is consistently positive across the redshift ranges

What is most interesting to note in the bias plots is that the SMBH bias essentially shifts from matching the sBBH to stellar bias at high redshifts to matching the sBBH to dark matter bias at low redshifts. This observation alone has implications for using sBBH clustering as a tracer of large-scale structure and is one we will return to in the discussion about differentiating formation channels.

\subsection{Observing times for Differentiating Biases}

Finally, we examine the evolution of the errors in the three biases for two distance bins at $1.2$ Mpc (top row of Figure ~\ref{fig:bias_time}) and $17.2$ Mpc (bottom rows of Fig ~\ref{fig:bias_time}). The former value was chosen as a rough boundary between the 1-halo and 2-halo terms for large-scale clustering and because it is on the lower end of what might be resolvable in the best-case scenarios for future detector networks. The latter was chosen because it is the largest distance bin from calculating the two-point correlation functions and relative clustering biases. Additionally, Table ~\ref{tab:yrs_table} summarizes relevant information from Figure ~\ref{fig:bias_time} about how many years of observing would be needed to distinguish biases, and thus, formation channels. 

For the case of $1.2$ Mpc separation, with the assumptions made in creating and analyzing this dataset, we are able to distinguish between the sBBH to stellar bias and both the sBBH to SMBH and sBBH to dark matter bias in less than ten years at all redshift ranges. In particular, redshift one stands out as a clear optimal redshift, as both channels are distinguishable with just one year of observing. We also show that for this smaller separation distance, the dark matter channel will likely be more easily testable than the AGN channel. For the case of $17.2$ Mpc, the sBBH to stellar and sBBH to SMBH biases continuously overlap out to nearly a decade of observations in nearly every case, leading us to believe that, at these higher separation distances, this method begins to lose its potency. $z = 1$ is the exception as the only redshift in which both channels could be distinguished in ten years of observation or less, and in the case of the stellar to AGN channel, just barely. It is also noteworthy that due to the different clustering evolution of dark matter over cosmic time, this may present an opportunity at higher redshifts to test the PBH channel through our clustering method.

\section{Discussion}
\label{sec:Discussion}

We have demonstrated that combining the Illustris and Illustris-sBBH datasets allows for a direct calculation of the two-point correlation functions and relative clustering biases of simulated ssBBH mergers. This represents the first application of the \textit{Illustris} framework to probe the spatial clustering of compact-object mergers and extends previous efforts linking sBBH populations to stellar and matter distributions. We find that the sBBH–stellar bias remains close to unity at both separation distances, though it is noteworthy that the sBBH-stellar bias is less than unity for both separation distances at $z = 3$. This deviation from unity at $z = 3$ indicates that merger events are not uniformly tracing their stellar hosts, and this redshift dependent sBBH-stellar clustering bias observation may provide a statistical window into how black hole binaries populate galaxies over cosmic time.

Assuming sBBHs form through isolated stellar binaries, we find that the relative clustering bias between sBBHs and either stellar particles or SMBHs becomes distinguishable in a next-generation facility such as CE at redshifts $z = 0.5$–5 on $\sim 1$~Mpc scales, though the observing time required varies strongly with redshift. At $z = 1$, the separation emerges after about one year of observation, while at $z = 3$–5, it requires up to a decade. In contrast, the sBBH–-dark matter bias diverges more rapidly, becoming separable within a year at $z \lesssim 3$. This behavior suggests that if primordial black holes contribute significantly to the observed merger population, clustering statistics would provide a more immediate constraint than for the stellar and AGN channels, which overlap more strongly in their spatial distributions.

A key trend across redshift is the evolution of the SMBH clustering bias—from following stellar clustering at $z \gtrsim 3$ to more closely matching dark matter at $z \lesssim 1$. This shift reflects the transition of SMBHs from residing in star-forming galaxies at early epochs to massive, quiescent halos at late times. Consequently, the redshift-dependent separation between stellar and SMBH clustering, rather than the total number of detections, primarily determines when the relative biases can be observationally distinguished.

Our work serves as a proof of concept and makes several simplifying assumptions. We neglect localization uncertainties in the synthetic gravitational-wave catalogs, approximate all SMBHs as AGN, assume primordial black holes follow the dark matter distribution, and omit the dynamical formation channel. These approximations enable a controlled first-order test of the clustering method but will be refined in future studies. Future work will include incorporating localization uncertainty for realistic network configurations, implementing a physical AGN prescription within Illustris, and extending this framework to next-generation simulations such as Illustris-TNG that capture baryonic feedback more accurately ~\citep{Nelson2019TNG}.

Finally, this approach can be extended to sources in space mission  LISA ~\citep{LISA_redpaper} and future lunar-based observatories such as Laser Interferometer Lunar Antenna (LILA) ~\citep{jani2025laserinterferometerlunarantenna, creighton2025fundamentalnoisegravitationalwavesensitivity} and Lunar Gravitational-Wave Antenna (LGWA) ~\citep{Harms_2021, Ajith_2025}, which will detect higher-mass black hole mergers at cosmological distances. If the clustering of intermediate-mass black holes differs from that of stellar-mass binaries, such comparisons could reveal distinct formation pathways across mass and frequency bands. Together, these results illustrate how gravitational-wave populations can serve as cosmological tracers, linking compact-object astrophysics with the large-scale structure of the Universe.

\section*{Software Acknowledgments}
This work made use of the following open-source software: NumPy for numerical computation ~\citep{NumPy}, Matplotlib for data visualization ~\citep{Matplotlib}, Corrfunc for high-performance correlation function calculations \citep{Corrfunc}, PyCBC for gravitational-wave data analysis ~\citep{Py_cbc}, and Astropy, a community-developed core Python package for astronomy ~\citep{astropy_2013, astropy_2018, astropy_2022}.

\section*{Acknowledgments}
WS, KRR, and KHB express their gratitude for funding for this work provided by the National Science Foundation (NSF 2125764). WS and KJ acknowledges support from Cornelius Vanderbilt Dean's Faculty Fellowship from Vanderbilt University. This work was conducted using the resources provided by the Vanderbilt Advanced Computing Center for Research and Education (ACCRE).  MM acknowledges financial support from the European Research Council for the ERC Consolidator grant DEMOBLACK, under contract no. 770017, and from the Deutsche Forschungsgemeinschaft (DFG, German Research Foundation) under Germany's Excellence Strategy EXC 2181/1 - 390900948 (the Heidelberg STRUCTURES Excellence Cluster).

\bibliography{Illustris_BBH}
\bibliographystyle{aasjournal}



\end{document}